 \documentclass[12pt,preprint]{aastex}

\usepackage[lowtilde]{url}
\usepackage{amsmath}
\usepackage{color}

\usepackage{graphicx}
\usepackage{epstopdf}

\newcommand {\Lya}    {Ly$\alpha$}   

\newcommand {\HI}     {\ion{H}{1}}      

\newcommand {\OVI}     {\ion{O}{6}}      
\newcommand {\OVII}    {\ion{O}{7}}
\newcommand {\OVIII}   {\ion{O}{8}}

\newcommand {\CIV}     {\ion{C}{4}}       

\newcommand {\SiIII}  {\ion{Si}{3}}

\newcommand {\FeII}   {\ion{Fe}{2}}

\newcommand {\kms}    {km~s$^{-1}$}

\newcommand {\cd}     {cm$^{-2}$}

\newcommand {\etal}   {et~al.} 

\begin{document}

\title{WHERE DO GALAXIES END? }

\author{J. Michael Shull }
\affil{ CASA, Department of Astrophysical \& Planetary Sciences, \\
University of Colorado, Boulder, CO 80309, USA}
\affil{and Institute of Astronomy, University of Cambridge, Cambridge, CB3~0HA, UK}  

\email{michael.shull@colorado.edu }  


\begin{abstract} 
Our current view of galaxies considers them as systems of stars and gas embedded in extended 
halos of dark matter, much of it formed by the infall of smaller systems at earlier times.  The true 
extent of a galaxy remains poorly determined, with the ``virial radius" ($R_{\rm vir}$) providing a 
characteristic separation between collapsed structures in dynamical equilibrium and external 
infalling matter.   Other physical estimates of the extent of gravitational influence include the 
gravitational radius, gas accretion radius, and  ``galactopause" arising from outflows that stall
at 100-200~kpc over a range of outflow parameters and confining gas pressures.  
Physical criteria are proposed to define bound structures, including a more realistic definition of 
$R_{\rm vir} (M_*, M_h, z_a)$ for stellar mass $M_*$ and halo mass $M_h$, half of which formed 
at ``assembly redshifts" ranging from  $z_a \approx 0.7-1.3$.   We estimate the extent of  bound 
gas and dark matter around $L^*$ galaxies to be $\sim200$~kpc.  The new virial radii, with mean
$\langle R_{\rm vir} \rangle \approx 200$~kpc, are 40-50\% smaller than values estimated in 
recent HST/COS detections of \HI\ and \OVI\ absorbers around galaxies.   In the new formalism, the 
Milky Way stellar mass, $\log M_* = 10.7 \pm 0.1$, would correspond to 
$R_{\rm vir} = 153^{+25}_{-16}$~kpc for half-mass halo assembly at $z_a =  1.06\pm0.03$.  The 
frequency per unit redshift of low-redshift \OVI\ absorption lines in QSO spectra suggests absorber 
sizes $\sim150$~kpc when related to intervening $0.1L^*$  galaxies.  This formalism is intended to 
clarify semantic differences arising from observations of extended gas in galactic halos, circumgalactic 
medium (CGM), and filaments of the intergalactic medium (IGM).  Astronomers should refer to {\it bound 
gas} in the galactic halo or CGM, and {\it unbound} gas at the CGM-IGM interface, on its way into the IGM.  

\end{abstract} 



\section{INTRODUCTION}

Over the last several decades, with new evidence, the objects we call ``galaxies" have become much larger.  
Extended dark-matter ``halos" were proposed to produce flat rotation curves at large radii in disk galaxies 
(Rubin \etal\ 1980), and a ``corona"  of hot interstellar gas at the galaxy's virial temperature was predicted by 
Spitzer (1956) to provide pressure confinement of high-latitude clouds. More recently, astronomers have
observed Galactic kinematic tracers (blue horizontal branch stars, globular clusters, satellite galaxies) to 
distances of 50-250 kpc, and X-ray absorption-line (\OVII) spectroscopy and stacked soft X-ray emission have 
provided evidence of large reservoirs of hot ionized gas in Milky Way halo (Miller \& Bregman 2013) and the 
outskirts of external galaxies (Soltan 2006; Anderson \etal\ 2013).  In ultraviolet spectroscopy, the Cosmic  Origins 
Spectrograph (COS) on the 
{\it Hubble Space Telescope} (HST) has recently detected extended (100-150 kpc) reservoirs of highly ionized 
oxygen (\OVI) around star-forming galaxies (Tumlinson \etal\  2011, 2013; Stocke \etal\ 2013) likely created by 
outflows of metal-enriched gas from star formation.  

Thus, our picture of galaxies has  evolved to a system of stars and gas embedded in an extended dark-matter 
halo, often associated with an even larger gaseous circumgalactic medium (CGM). Galaxies must also be viewed 
in a cosmological context, in which most of the baryonic matter in the universe is unseen (Persic \& Salucci 1992) 
and likely distributed through the intergalactic medium (IGM) in a ``cosmic web" shaped by dark-matter structure and
inefficient galaxy formation (Cen \& Ostriker 1999; Dav\'e \etal\ 1999; Smith \etal\ 2011).  Observations and modeling 
(Shull \etal\ 2012) suggest that 60-80\% of the cosmological baryons reside in the low-redshift, multi-phase IGM, 
with perhaps 20\% in collapsed form (galaxies, groups, clusters).  
Although these baryon fractions have some uncertainty, current research is devoted to understanding the 
physical conditions, spatial extent, and evolution of the gas at distances of 100~kpc to a few Mpc from galaxies.  
At low redshift ($z \leq 0.4$) the influence of galactic winds and metal injection out to $\sim1$~Mpc has been 
inferred from the association of QSO absorption systems of \HI\ and \OVI\ with nearby galaxies
(Penton \etal\ 2002;  Prochaska \etal\ 2011; Stocke \etal\ 2006, 2013).
Galactic outflows have been detected in absorption toward intermediate-redshift galaxies 
(Steidel \etal\ 2004; Martin \etal\ 2012; Tripp \etal\ 2011), while at higher redshifts, $z \approx 2.0-2.8$,
Rudie \etal\ (2012) found an enhancement of circumgalactic ($\leq300$~kpc)  \HI\ absorbers in a sample
of 886 star-forming galaxies probed by 15 background QSOs.  Thus, the connection between galaxies and 
extended absorption systems seems secure.

Perhaps because of the recent nature of these discoveries, a semantic problem has arisen regarding the proposed 
structures:  halo, CGM, IGM.  When does gas cease to be circumgalactic and become intergalactic?   Are the edges
of galaxies defined by gravity or gas outflows?  
Are quasar absorption lines the extended halos of intervening galaxies (Bahcall \& Spitzer 1969) or filaments 
of intergalactic gas (Sargent \etal\ 1980)?   Similarly, the phrase ``circumgalactic medium" appears to have 
replaced the concept of a ``gaseous halo" or ``galactic corona"  of hot interstellar gas at the galaxy's virial temperature.   
Ionized gas with high covering factor has been detected above the Galactic disk in UV absorption-line surveys in 
metal ions such as \OVI\ (Sembach \etal\  2003) and \SiIII\ (Shull \etal\ 2009;  Lehner \& Howk 2011) and in soft X-ray
absorption lines of  \OVII\ or \OVIII\  at $z \sim 0$  (Nicastro \etal\ 2002; McKernan \etal\  2005; Wang \etal\ 2005).    The 
UV absorbers appear to come primarily from gas within 2-10~kpc of the disk plane, elevated by supernovae and star 
formation in the disk.  The Galactic X-ray absorption suggests hot gas at $T \approx 10^{6.3\pm0.2}$~K, 
but its radial  extent is controversial.   It may come from a 50-kpc halo (Anderson \& Bregman 2010; Gupta \etal\ 2012)
although  \OVII\ absorption toward background AGN (Fang \etal\ 2006; Hagihara \etal\  2010) and  X-ray binaries 
(Yao \& Wang 2005; Hagihara \etal\ 2011) suggests that much of the absorption comes within several kpc of the disk.  
The disk model is consistent with both  X-ray observations and total mass considerations (Collins \etal\ 2005; Fang 
\etal\ 2006).  

The intent of this paper is to improve the definition of galaxy halos as regions of strong gravitational influence, using
dynamical principles and observational constraints.  Section~2 discusses physical measures of the spatial extent 
of large ($\sim L^*$) galaxies including the Milky Way and Andromeda.  We discuss the somewhat arbitrary and 
occasionally misused definition of  ``virial radius".  Somewhat better defined are the  ``gravitational radius", 
$GM^2/\vert W \vert$, derived from galactic mass and  potential energy, the gravitational sphere of influence, 
accretion radius, and tidal radius.  Section 2.3 develops a physically realistic definition of $R_{\rm vir} (M_h, z_a)$ 
for halos of mass $M_h$, assembled primarily at redshifts $z_a \approx 0.7-1.3$ with further mass accretion down to 
the present epoch.  These new virial radii are typically 50-60\% the sizes used to analyze hot halo gas with HST/COS.  
Section 3 discusses four estimates of galaxy extent:  
(1) recent kinematical studies of the Milky Way and Andromeda (M31); 
(2) the ``galactopause" where outflow ram pressure balances  thermal pressure of the CGM; 
(3) QSO absorber cross sections derived from metal absorption-line frequency in redshift; and
(4) virial radii and halo masses obtained from galaxy abundance-matching.    We also discuss recent estimates of the 
mass and size of the Milky Way and Andromeda halos from kinematics of stars, galactic satellites, and the Local Group.   
Most of our estimates suggest a smaller spatial extent ($\sim200$~kpc) for galaxies of mass $\sim 10^{12}~M_{\odot}$, 
comparable to the Milky Way and M31.  Section~4 concludes with applications of the new definition of virial radius to 
observations of extended gas around galaxies made with HST/COS, and to recent mass and size measurements for 
the Milky Way and M31.  

\newpage

\section{DEFINITIONS OF GALAXY EXTENT }  

This section begins with an overview of the cosmological context, in which galaxies form from
density perturbations that turn around from the Hubble expansion background and collapse into 
dynamical (virial) equilibrium, but with continued infall of dark matter and gas.  The environment
of these structures will be influenced by ``feedback" from star formation within galaxies, through
outflows that escape to the IGM or become stalled and return to the galaxy.   Both infall and outflow 
processes depend on the gravitational attraction of the galaxy and its halo, as measured by parameters 
such as escape velocity, accretion radius, and tidal radius.  Below, these measures of gravitational 
influence are explored, leading to a practical definition of virial radius, consistent with collapse in 
the past ($z \approx 1$) with continued mass infall to the present.  

The classic overdensity $\Delta_{\rm vir} = 18 \pi^2 \approx 178$ (often rounded to 200) was originally
derived analytically for a spherical top-hat density perturbation in an Einstein-deSitter universe.  The 
mean background density, $\bar{\rho}(t) \propto t^{-2}$, following the behavior of the expansion parameter 
$a(t) \propto t^{2/3}$.  The perturbation begins its collapse at a ``turn-around" density, 
$\bar{\rho}_{\rm turn} = (9 \pi^2/16) \bar{\rho}_m \approx 5.55 {\bar{\rho}}_m$, at a redshift $z_{\rm turn}$.  
For example, if  all the matter in the Local Group were associated with a spherically distributed mass 
$M_h$, the proper size of the mass perturbation at turnaround would be
\begin{equation}
   R_{\rm turn} \approx \left[ \frac {M_h} { (4 \pi/3) \, (5.55) \,  {\bar{\rho}}_m(z) } \right] ^{1/3}
          \approx (1.2~{\rm Mpc}) \left( \frac {M_{\rm LG}} {5\times10^{12}~M_{\odot} } \right) ^{1/3}  
          \left[ \frac {1.5}{ 1+z_{\rm turn} } \right]   \; .
\end{equation}
Here, we have scaled to the local group mass, $M_{\rm LG} \approx 5\times10^{12}~M_{\odot}$ at a
turnaround redshift $z_{\rm turn} \approx 0.5$, corresponding to the recently measured dynamical 
history of the Milky Way and Andromeda system (van der Marel \etal\ 2012).   These galaxies
turned around from the Hubble flow even earlier, perhaps at $z \approx 2-3$, and had correspondingly 
smaller radii, probably $0.25-0.30$~Mpc.  The fact that a given galaxy halo is assembled over 
a range of redshifts suggests an improved definition of virial radius, $R_{\rm vir} (M_h, z, z_a)$, for a 
galaxy observed at redshift $z$ but  assembled primarily at redshift $z_a > z$, when half its halo 
mass had collapsed.  This change in definition increases the background matter density, $\rho_m(z_a)$, 
and makes the virial radii smaller.    We return to this new formalism in Sections 2.3 and 3.4.

\newpage

\subsection{Gravitational Radius}  
  
For extended self-gravitating systems without a sharp boundary, it is useful to define a characteristic 
``gravitational radius" (Binney \& Tremaine 2008) as $r_g = GM^2/ \vert W \vert$, where $W$ is the 
gravitational potential energy,
\begin{equation}
    W = \frac{1}{2} \int  \,   d^3 \mathbf{x} \, \rho(\mathbf{x}) \, \Phi(\mathbf{x})  
        =  -4 \pi G \int_{0}^{\infty} \rho(r) \, M(r) \, r \, dr   \;  .  
 \end{equation}
 The gravitational radius also bears a close relation to the ``half-light radius", $r_h$, defined by 
 Spitzer (1969) for spherical stellar systems.  For many mass distributions,  $r_h/r_g = 0.4-0.5$, 
so that  $r_h  \approx 0.45 (GM^2/ \vert W \vert)$ and $\langle v^2 \rangle \approx GM/r_g$.  
 For example, a homogeneous sphere of mass $M$ and radius $R$ has $W = -3GM^2/5R$ and 
$r_g = 5R/3$.   For the density distribution of  Plummer's model,  
$\rho(r) = (3M/4 \pi b^3)[1 + (r/b)^2]^{-5/2}$, $W = -(3 \pi GM^2 / 32 b)$ and $r_g \approx 3.40 b$,
where $b$ is the Plummer scale length.  
  
The structure of collapsed halos has been well studied through cosmological N-body simulations
(Navarro, Frenk, \& White 1997;  Springel \etal\ 2005;  Klypin \etal\ 2011).  These numerical
experiments  show that dark-matter halos collapse into structures with  cuspy cores and extended 
halos.  The collapsed structures in these  simulations have been fitted to various radial profiles, 
such as NFW  (Navarro, Frenk, \& White 1997), for which the density, potential, and enclosed mass are,
\begin{eqnarray}   
  \rho(r)  &=&   \frac {\rho_0} {(r / r_s )  \left[ 1 +  (r / r_s ) \right] ^2  }    \\   
   \Phi(r) &=& - \left( 4 \pi G \rho_0 r_s^2 \right) \frac { \ln (1 + r/r_s)} {r/r_s}   \\
       M(r) &=& \left( 4 \pi \rho_0 r_s^3 \right) \left[ \ln \left( 1 +  \frac{r}{r_s} \right) - \frac {r / r_s}{1+ r / r_s} \right] \; .
\end{eqnarray}
Here $r_s$ is a characteristic radius defined by the break in slope and related to the virial radius through 
$r_s = R_{\rm vir}/c$, with the ``concentration parameter", $c$, typically between  5--20.  The mass 
enclosed with the virial radius is written as
\begin{equation}
   M_{\rm vir} \equiv M(R_{\rm vir}) = \left( 4 \pi \rho_0 r_s^3 \right) \left[  \ln (1+c) - \frac {c}{1+c} \right]    \;  ,
\end{equation}  
where $M_{\rm vir}$ and $c$ are found to be correlated in cosmological simulations (Navarro \etal\ 1997;
Prada \etal\ 2012).

Increasing the concentration, $c$, adds mass and extent to the halo.  For example, changing $c$ from 
5 to 20 doubles the enclosed mass, $M_{\rm vir}(c = 20) /  M_{\rm vir}(c = 5) \approx 2.18$.  The circular 
velocity, $V_c^2(r) = G M(r)/r$, reaches a maximum $V_{\rm max}$ at radius 
$r = 2.1626 r_s = 2.1626 R_{\rm vir}/c$, and equal to
\begin{equation}
   V_{\rm max}^2  = \left( \frac {GM_{\rm vir}}{R_{\rm vir}} \right) 
          \frac {0.2162 \, c} { \left[  \ln (1+c) - \frac {c}{1+c} \right] } \; .
\end{equation}
For the NFW model,  the mass diverges logarithmically with radius, and the gravitational radius is 
formally undefined without a cut-off.  By integrating Equation (2) out to the virial radius, 
$R_{\rm vir} = c r_s$, we can define a finite NFW potential energy and gravitational radius,
\begin{eqnarray}
   W_{\rm vir} ^{\rm (NFW)}  &=& - \left( \frac  {GM_{\rm vir}^2} {2 \, r _s} \right) 
    \frac { \left[ 1 - \frac {\ln (1+c)} {(1+c)}  - \frac {1}{(1+c)} \right] }
   { \left[  \ln (1+c) - \frac {c}{1+c} \right]^2  }    \\
 r_g ^{\rm (NFW)}             &=&  GM_{\rm vir}^2 / \vert W_{\rm vir} \vert = 
         (2 r_s) \;  \frac {  \left[ \ln (1+c) - \frac {c}{1+c} \right]^2  } 
          {  \left[ 1 - \frac {\ln (1+c)} {(1+c)}  - \frac {1}{(1+c)} \right] }  \; .
\end{eqnarray}   
This NFW gravitational radius ranges from $r_g = 3.44r_s = 0.69 R_{\rm vir}$ (for $c = 5$) to 
$r_g = 10.8 r_s = 0.54R_{\rm vir}$ (for $c = 20$).  Table 1 summarizes the ratio of gravitational
radius to characteristic scale lengths for several commonly used density distributions.

\subsection{Gravitational Radius of Influence and Accretion Radius}

Another observationally useful characteristic radius is the gravitational  ``radius of influence" (Ferrarese 
\& Ford 2005; Binney \& Tremaine 2008)  defined as the radius at which the Kepler velocity around a central 
mass $M$ equals the transverse velocity dispersion, $\sigma_{\parallel}$, of surrounding stars,
\begin{equation} 
   R_{\rm infl} = \frac {GM} {\sigma_{\parallel}^2} = (191~{\rm kpc}) M_{12} \, \sigma_{150}^{-2}  \; .
\end{equation} 
In the galactic context of this paper, we assume that $M$ is the total mass of the galaxy within
radius $R_{\rm infl}$, scaled as $M = (10^{12}~M_{\odot}) M_{12}$.  The gravitational influence of
this mass can traced by luminous halo stars or galaxy satellites with dispersion
$\sigma_{\parallel}  = (150~{\rm km~s}^{-1}) \sigma_{150}$.  The dynamical dependence of 
$\sigma$ (observed in the stellar component) on $M$ will soften the linear dependence of 
$R_{\rm infl}(M)$.   In the Milky Way, radial velocity dispersions, $\sigma_r \approx 110$~\kms, of 
different halo populations have been measured out to 80~kpc (Xue \etal\ 2008; Gnedin \etal\ 2010).  
From radial velocities of blue horizontal branch stars between 100-150 kpc,  Deason \etal\ (2012) 
find even lower dispersions, $\sigma_r = 50-70$~\kms, although values for satellites are somewhat larger.   
Thus, depending on assumptions about the ratio, $\sigma_{\parallel} / \sigma_r$,  the value of $R_{\rm infl}$
for the Milky Way halo may be somewhat larger than 200~kpc.    

This formula is analogous to that used in Bondi accretion of gas with isothermal sound speed 
$c_s$, for which one defines an ``accretion radius"
\begin{equation}
   R_{\rm accr} = \frac {2GM} {c_s^2} = (202~{\rm kpc}) M_{12} \, T_{6.5}^{-1} \; .
\end{equation}
Here again, we have scaled to a $10^{12}~M_{\odot}$ galaxy surrounded by a gaseous halo at 
its virial temperature $T =  (10^{6.5}~{\rm K}) T_{6.5}$ with 
$c_s = (kT/ \mu)^{1/2} = (207~{\rm km~s}^{-1}) T_{6.5}^{1/2}$ for mean atomic weight $0.61 m_H$ 
appropriate for  a fully ionized plasma with $n_{\rm He}/n_{\rm H} = 0.1$ by number.  
The fact that these two raidii are similar is a consequence of the chosen gas temperature and
stellar velocity dispersion, which are expected to be in near equilibrium.

\subsection{Virial Radius} 

In standard cosmological terminology, the virial radius, $R_{\rm vir}$, is defined such that the 
mean density of a halo of mass $M_h$ equals the ``virial overdensity" (variously denoted as
$\Delta_{\rm vir}$ or $\Delta_c$) times the mean density of the universe.  Unfortunately, some 
confusion has arisen in the proper value of $\Delta_{\rm vir}$, much of it over convention:  
whether the overdensity is expressed relative to the critical (closure) density, 
$\rho_{\rm cr} = (3 H_0^2 / 8 \pi G)$, or to the matter density, $\rho_m = \Omega_m \rho_{\rm cr}$.  
Both conventions are used in the literature, but if one is not careful in using a consistent definition 
of $\Delta_{\rm vir}$, the estimated virial radius will be too small by a factor 
$\Omega_m^{1/3} \approx 0.67$ and the mean matter density of the halo too large by  
$\Omega_m^{-1}  \approx 3.3$.   There are other misapplications of the virial radius because
the initial collapse to dynamical equilibrium occurred in the past, when $\rho_m$ was higher,
followed by additional infall during galaxy assembly.  For these reasons, we develop a 
physically motivated definition of $R_{\rm vir}(M_h, z, z_a)$  for a halo of mass $M_h$,
observed at redshift $z$ but assembled at an earlier epoch, $z_a > z$.  The redshift of halo assembly 
$z_a$ is taken as the epoch of major virialization ($z_{\rm vir}$) when half the halo mass had collapsed, 
a useful convention introduced by Tacchella, Trenti, \& Carillo (2013) hereafter denoted TTC13.  
  
The basic concept of virialization follows from the assumption of conservation of kinetic energy and 
potential energy during collapse ($T + W = {\rm constant}$) together with the virial theorem ($2T = -W$).  
The gravitating system is assumed to collapse to half its turnaround radius at twice the turnaround time 
for cycloidal collapse.   The collapsed system has then increased in mean density by a factor of 8 since 
turnaround.   In an Einstein-deSitter universe ($\Omega_0 = 1$),  the expanding background density, 
 $\bar{\rho}(t) \propto t^{-2}$, has dropped by a factor of 4 for an expansion parameter $a(t) \propto t^{2/3}$.
Thus, one obtains the classic virial overdensity, 
$\bar{\rho}_{\rm vir} = 32  {\bar \rho}_{\rm turn} = (18 \pi^2) {\bar{\rho}}$, where  $\bar{\rho}_{\rm turn}$ and 
$\bar{\rho}_{\rm vir}$ are evaluated at the times of turnaround and virialization, respectively. 
Cole \& Lacey (1996) explored the structure of halos in N-body simulations of clustering in 
an $\Omega_0 = 1$ universe.  They found that the radius, $r_{178}$ enclosing a mean overdensity of 178, 
accurately separated the virialized halo interior, in approximate dynamic equilibrium, from the exterior
where matter was still falling in.  This property was noted by Binney \& Tremaine (2008), who recommended 
using $\Delta_{\rm vir} = 200$.  

The situation changes in a flat $\Lambda$CDM cosmology ($\Omega_m + \Omega_{\Lambda} = 1$) where 
both the expanding background and collapse dynamics differ from the $\Omega_0 = 1$ case.   Eke \etal\ (1996) 
studied virialization dynamics for flat models with $\Omega_m < 1$, but they defined overdensity relative to 
{\it closure} density.  For values of $\Omega_m \approx 0.3$ at $z \approx 0$, they found 
$\Delta_c \equiv {\bar \rho}_{\rm vir} / \rho_{\rm cr} \approx 100$.  In Figure 1 of their paper, they plotted the 
dependence of $\Delta_c$ on $\Omega_m$, which serves as a proxy for higher redshift, since $\Omega_m(z)$
approaches 1 at the higher redshifts when most systems collapsed.  
Bryan \& Norman (1998) provided an approximate fit,  $\Delta_{\rm vir} (z) = [18 \pi^2 + 82x - 39 x^2] $ where 
$x = (\Omega_m - 1)$.   This expression must be divided by $\Omega_m(z)$ to be consistent with the
convention that overdensity is relative to {\it matter} density.  We define $R_{\rm vir}$ through the relation
$M_h = (4 \pi R_{\rm vir}^3 /3) \Delta_{\rm vir} (z_{\rm vir} ) {\bar  \rho}_m (z_{\rm vir})$, where the
mean matter density at the virialization (or assembly) redshift is
${\bar \rho}_m (z_{\rm vir}) = \Omega_{m,0} (1+z_{\rm vir})^3 \rho_{\rm cr,0}$, and the parameters
$ \Omega_{m,0}$ and $\rho_{\rm cr,0}$ are defined at the current epoch ($z = 0$).  Further discussion is given 
by Klypin \etal\ (2011) and van der Marel \etal\ (2012), both of whom follow the convention that $\Delta_{\rm vir}$
is relative to the mean {\it matter} density.  Their expressions are scaled to a product, 
$\Omega_m  \, \Delta_{\rm vir}  \approx 97.2$,  evaluated at $z = 0$ for which $\Delta_{\rm vir} = 360$ 
and $\Omega_{m,0} = 0.27$.

However, many papers in the literature do not account for the fact that virial collapse occurred at earlier times,
at redshift $z_{\rm vir}$ when $ \Omega_{m}(z) >  \Omega_{m,0}$.  For a flat $\Lambda$CDM universe with
$\Omega_{m,0} =  0.27-0.30$ at $z \approx 0$, one finds $\Delta_{\rm vir} \approx 350\pm10$ (relative to 
$\rho_m$) and a product $\Omega_m \Delta_{\rm vir} \approx 100$.   For the more typical collapse 
at higher redshifts,  the product $\Omega_m \, \Delta_{\rm vir} \approx 200$, appropriate for the decrease in 
$\Delta_{\rm vir}(z)$ as $\Omega_m(z) \rightarrow 1$.  This overdensity of 200 is slightly above the classical 
value $\Delta_{\rm vir} \approx 178$ for dynamical reasons discussed by Binney \& Tremaine (2008).   
Consider a galaxy observed at redshift  $z$,  associated with halo mass $M_h = (10^{12} M_{\odot}) M_{12}$
and assembled (at $z_a$) with half its current mass.   The virialization redshift $z_{\rm vir}$ is taken to be $z_a$, 
determined from standard cosmological collapse criteria (Lacey \& Cole 1993;  Sheth \& Tormen 1999)
and computed as in Trenti, Perna, \& Tacchella (2013).   Table 2 lists the assembly redshifts,  which range from 
$z_a \approx 1.35$ (at $M_h = 10^{11}~M_{\odot}$) to $z_a \approx 0.81$ (at $M_h = 10^{14}~M_{\odot}$) for 
various galaxies observed at $z \approx 0.2-0.3$ with stellar masses $M_*$.    Based on these new criteria for 
galaxy collapse and assembly, the virial radius is:
\begin{equation}
   R_{\rm vir} (M_h, z_a) = (206~{\rm kpc}) \, h_{70}^{-2/3} \, M_{12}^{1/3}  \left[  \frac 
       {\Omega_m (z_{\rm a})  \, \Delta_{\rm vir} (z_{\rm a})}  {200} \right] ^{-1/3} 
        \left( 1+z_a \right)^{-1}  \;  .
\end{equation}
Although the relation between $z_a(M_h)$ exhibits a small anti-correlation (Table 2), it does contain some  
numerical scatter which should be kept in mind. 
This expression differs from often-used formulae by the over-density scaling and by the factor $(1+z_{\rm a})^{-1}$, 
reflecting the fact that most galaxies  underwent virialization in the past, not at $z = 0$.   After this initial (half-mass) 
assembly, their proper size changes gradually because of continued mass infall into the halo.   To lowest order, this 
slight growth  in $R_{\rm vir}$ ($2^{1/3} \approx 1.26$) is captured in Eq.\ (12) by using the {\it current} halo mass ($M_h$).  
Second-order effects of late infall or adiabatic compression may occur in the radial mass distribution, which in the 
NFW formalism changes the scale parameter $r_s$ and concentration $c$.   Some of the infalling matter may not 
attain virial equilibrium, as the orbital timescale is quite long,
$t_{\rm orb} = 2 \pi R_{\rm vir}^{3/2}/(GM_h)^{1/2} \approx (8.64~{\rm Gyr}) (1+z_a)^{-3/2}$.
This formalism assumes spherical infall and a constant density profile; both approximations change in more realistic 
collapse scenarios.  For example, after accounting for central condensation of the collapsed structure, 
Rubin \& Loeb (2013) find a reduced virial overdensity.

\subsection{Escaping Stars Injected into the Halo}  

In addition to infall, galaxy halos can be probed by fast stars injected radially outward.  
We therefore discuss the radius at which the fastest galactic stars reach maximum apogalactic distance  in a 
dark matter halo potential.  As a simple model that illustrates the basic effect, we assume an isothermal sphere 
distribution in density, $\rho(r) = (\sigma^2 / 2 \pi G r^2)$ and enclosed mass, $M(r) = (2 \sigma^2/G)r$ leading to 
a constant circular velocity $V_c^2 = G M(r)/r = 2 \sigma^2$.   For this density and mass distribution, the 
equation of motion is  $\ddot{r} = -G M(r)/r^2$, for a star launched with radial velocity $\dot{r}  = v_0$ from radius 
$r = r_0$, leading to the first integral,
\begin{equation}
   \frac {1}{2} \left[ v^2 - v_0^2 \right] = - (2 \sigma^2) \ln (r / r_0) \;  .
\end{equation}
If the star coasts to rest ($v = 0$ at $r = r_{\rm max}$) one has $v_0^2 = 4 \sigma^2 \ln (r_{\rm max}/r_0)$
and thus
\begin{equation}
   r_{\rm max} = r_0 \exp \left[  (v_0 / 2 \sigma)^2 \right]   \;   .
\end{equation}
Equation (14) illustrates the sensitivity of $r_{\max}$ to launching conditions ($v_0$ and $r_0$) and halo mass 
(velocity dispersion $\sigma$).   For example, a star with $v_0 = 500$~\kms\ injected radially from
$r_0 = 8.3$~kpc into a halo with $\sigma = 200$~\kms\ would have a maximum extent of just 40 kpc.
The stellar ejection distance doubles ($r_{\rm max} \approx 79$~kpc) if the launch velocity $v_0 = 600$~\kms.  
These distances are not surprising for velocities comparable to the escape speed from the solar circle within the 
Milky Way ($r_0 = 8.28$~kpc), recently estimated from RAVE stellar survey data to be 
$V_{\rm esc} = 533^{+54}_{-41}$~\kms\  (Piffl \etal\  2014).  

Repeating the above calculation for the maximum radius in an NFW mass distribution and potential
(Equations 3, 4, 5) we find,
\begin{equation}
   v_0^2 = \left( \frac {G M_{\rm vir}} {R_{\rm vir}} \right) \frac {2c} { \left[ \ln (1+c) - \frac {c}{1+c} \right] } 
      \left[  \frac {\ln \left( 1 + r_0 / r_s \right) } {r_0 / r_s}  -
                 \frac {\ln \left( 1 + r_{\rm max} / r_s \right) } {r_{\rm max} / r_s}  \right]   \; .
\end{equation}
This equation can be solved for the velocity $v_0$ required for a star to reach radius $r_{\rm max}$
from initial radius $r_0$.  Combined with equation (7) for the maximum circular velocity, $V_{\rm max}$,
in an NFW halo, we can write the velocity $v_0$ needed to escape to $r \rightarrow \infty$ as
\begin{equation}
  v_0 = V_{\rm max}  \left[ \frac {2} {0.2162} \;  \frac { \ln \left( 1 + r_0 / r_s \right)} { (r_0 / r_s)} \right] ^{1/2} \; ,
\end{equation}
independent of $c$.  If  we take $r_0 = 2.1626 \, r_s$ as the radius at $V_{\rm max}$, we find 
$v_0 = 2.219 V_{\rm max}$.  This agrees reasonably well with the Milky Way values, if we assume
$V_{\rm max} \approx 240$~\kms\ and obtain an escape velocity $v_0 \approx 533$~\kms.  

A related size estimate comes from the hypothetical radius, $r_{\rm flat}$, to which one must extend an 
isothermal density distribution to maintain a constant (flat) circular velocity,
\begin{equation}
   r_{\rm flat} = r_0 \exp \left[ \frac {V_{\rm esc}^2}{2 V_{\rm c}^2} - 1 \right]  \;   .
\end{equation}  
Of course, rotation curves do not remain flat at large radii, but $r_{\rm flat}$ provides a lower limit on 
the extent of the isothermal model, as noted by Smith \etal\ (2007).
For $V_{\rm esc} = 533^{+54}_{-41}$~\kms\ and $V_{\rm c} = 220$~\kms, one finds a truncation radius 
$r_{\rm flat} \approx 60$~kpc for $r_0 = 8.3$~kpc, with a large range (40 to 110 kpc) arising from
uncertainties in determining $V_{\rm esc}$ and $V_c$.   At larger radii, the circular velocity, 
$V_c(r)$ likely declines, particularly if the halo density distribution falls off as $r^{-3}$ as in the NFW 
model.   The total (virial) masses of galaxies include substantial mass beyond 50-80 kpc,
particularly for halos with $c \geq 10$.  Extra mass accretion may account for the discrepancy between low 
and high values for the Milky Way mass estimates.  It is also consistent with the physically motivated 
revision to the method in which the virial radius is calculated at the ``galaxy assembly time" at
redshift $z_a > z$ (Section 2.3).  We return to these observational issues in Section 3.4. 

With the current state of observations, these kinematic constraints suggest halo sizes of 150-200 kpc
for $L^*$ galaxies.  
Future observations of rotation curves, $V_c(r)$, and more distant kinematic  tracers (blue horizontal branch 
stars, hypervelocity stars, globular clusters, and satellites) will be of great help.  The RAVE survey data will 
be greatly improved by the {\it Gaia} mission, leading to better determinations of escape velocity from the 
solar circle, and later with LSST, through studies of galaxy satellites, BHB stars, and RR-Lyrae stars.

\section{OBSERVATIONS OF GALAXY EXTENTS}  

\subsection{The Milky Way and Andromeda System} 

Despite their proximity, the Milky Way (MW) and Andromeda (M31) galaxies remain surprisingly
uncertain in their inferred mass and extent.   Recent MW mass estimates range from just under
$10^{12}~M_{\odot}$ (Smith \etal\ 2007; Xue \etal\ 2008; Gnedin \etal\ 2010; Deason \etal\ 2012)
to values between $(1.2-2.0)\times10^{12}~M_{\odot}$ (Boylan-Kolchin \etal\ 2013; Nesti \& Salucci 
2013).  Some of the lower masses are inferred for stellar tracer distances well within $R_{\rm vir}$.  
For example, Gnedin et al. (2010) use hypervelocity stars at distances beyond 25~kpc to find 
$M(\leq 80~{\rm kpc})  = 6.9^{+3.0}_{-1.2} \times 10^{11}~M_{\odot}$.  Xue \etal\ (2008) use 2400 
blue horizontal branch stars to estimate 
$M(\leq 60~{\rm kpc})  = (4.0 \pm 0.7) \times 10^{11}~M_{\odot}$.  However, they extrapolate to find a
virial  mass $M_{\rm vir} = (1.0-1.2)  \times 10^{12}~M_{\odot}$ for two models of the circular velocity 
curve.  Busha \etal\ (2011) use LMC and SMC kinematics to infer 
$M_{\rm MW} = (1.2^{+0.7}_{-0.4}) \times 10^{12}~M_{\odot}$.

Boylan-Kolchin \etal\ (2013) provide a thorough discussion of the previous MW  mass measurements,
both large and small, but they base their  mass  on the high probability that the Leo~I satellite is 
gravitationally bound.  For a Leo~I distance ($D = 261\pm13$~kpc) and Galactocentric space velocity 
(200~\kms), they find a median Milky Way virial mass $M_{\rm MW} = 1.6 \times 10^{12}~M_{\odot}$ 
with an asymmetric 90\% confidence interval of $[1.0-2.4]\times10^{12}~M_{\odot}$ and 
$R_{\rm vir} \approx 300$~kpc for the median mass.  Nesti \& Salucci (2013) use dark-matter halo 
models to estimate $M_{\rm vir} \approx (1.1-1.5) \times 10^{12}~M_{\odot}$ 
but with substantial uncertainties for both Burkert and NFW density profiles.  Piffl \etal\  (2014) 
use the inferred escape speed at the solar circle, $533^{+54}_{-41}$~\kms\  from the RAVE survey, to 
estimate $M_{\rm MW} = (1.6^{+0.5}_{-0.4}) \times 10^{12}~M_{\odot}$ and a virial radius
$R_{200} = 225\pm20$~kpc.  Watkins \etal\ (2010) find 
$M_{\rm MW} = (1.4\pm0.3)\times10^{12}~M_{\odot}$ within 300~kpc from 26 satellite galaxies, and
McMillan (2011) proposes a mass model with $M_{\rm vir} = (1.26\pm0.24) \times 10^{12}~M_{\odot}$.  
Based on the recent dynamical estimates out to larger distances, we conclude that the Milky Way 
virial mass is likely close to $1.6 \times 10^{12}~M_{\odot}$. 

Mass estimates for the Andromeda (M31) galaxy include  $(1.4\pm0.4)\times10^{12}~M_{\odot}$ 
within 300 kpc from 23 satellite galaxies (Watkins \etal\ 2010) and $(1.2-1.5) \times10^{12}~M_{\odot}$ 
within 200 kpc from kinematics of 53 globular clusters (Veljanoski \etal\ 2013).   Tamm \etal\ (2012) 
used photometry and the M31 rotation curve, together with several mass distributions, to find a dark-matter
halo mass $M_{200} = (0.8-1.1) \times 10^{12}~M_{\odot}$, stellar mass 
$M_* = (1.0-1.5) \times 10^{11}~M_{\odot}$, and virial radius 189-213~kpc.  Fardal \etal\ (2013) used 
Bayesian simulation models of the giant southern stream around Andromeda to derive 
$\log (M_{\rm M31} / M_{\odot}) = 12.3\pm0.1$.  They note that this mass, and those recently inferred for the 
Milky Way, alleviate the tension between virial mass estimates and the Local Group timing mass, recently 
estimated at $(4.93\pm1.63 )\times10^{12}~M_{\odot}$ (van der Marel \etal\  2012).    Because both the
Milky Way and Andromeda halos have likely accreted substantial mass beyond their observable radii 
(60-80 kpc), their virial masses are taken here to be $M_{\rm MW} = (1.6 \pm 0.4) \times10^{12}~M_{\odot}$ and 
$M_{\rm M31} = (1.8 \pm 0.5) \times10^{12}~M_{\odot}$, with a sum, $(3.4 \pm 0.6) \times 10^{12}~M_{\odot}$, 
close to the inferred timing mass of the Local Group.  Simulations suggest that the remaining mass may 
consist of ``diffuse dark matter" produced by the destruction of sub-halos (Gao \etal\ 2004).  Cosmic scatter 
could reduce the Local Group mass from the value above, as has been suggested by
van der  Marel \etal\ (2012) and Gonzalez, Kravtsov, \& Gnedin (2013).  

Another measure of gravitational influence is from tidal effects between nearby galaxies.  
We define the characteristic tidal distance in terms of masses and densities of the two galaxies, 
treated as deformable fluid spheres  (Chandrasekhar 1963) of masses ($M_M$, $M_m$) and 
radii ($R_M$, $R_m$), 
\begin{equation}
   d_{\rm tide} = 2.44 \left( \frac {\rho_M} {\rho_m} \right)^{1/3} R_M 
                = 2.44 \left( \frac {M_M} {M_m} \right)^{1/3} R_m   \; .  
\end{equation}
In the case of the Milky Way (MW) and Andromeda (M31) system, internal  tidal influences are seen in the 
form of large-scale streams of stars (Fardal \etal\ 2013; Ibata \etal\ 2014), but no obvious tails or bridges 
produced by external tides are detected.  Ibata et al. (2014) find a smooth stellar-light halo in M31 out 
to 150 kpc, with internal streams from tidally stripped satellites.  Usually the Roche formula is
applied to situations with $M_M \gg M_m$.  However, in the case of the MW-M31 system, we assume 
$M_M \approx M_m$ to estimate that $d_{\rm tide} \approx 2.44 R_M$ for each galaxy.  Here, $R_M$ 
can be regarded as the effective radius of either galaxy, each of mass $(1-2)\times 10^{12}~M_{\odot}$.  
If the distance between the two galaxy centers is $D = 770 \pm 40$~kpc (van der Marel \etal\ 2012),
the absence of strong mutual tidal effects limits their tidal radius to $d_{\rm tide} < D/2$.   
Equation (18) then implies that most of the stellar and gaseous matter in the two galaxies  is confined within  
radius $R_M < (D/2)/2.44 \approx 160$~kpc. The fluid approximations in making this estimate are large, 
and more accurate, time-dependent models  and searches for external tidal effects would be helpful to set a 
better limit.

\subsection{Gaseous Outflows}  

Many galaxies undergoing active star formation have been spectroscopically observed at $1.4 \leq z \leq 2.5$
(Steidel \etal\ 2004) to have large-scale outflows with bulk velocities of 200-300 \kms. For a sample of over 
200 star-forming galaxies at redshifts $0.4 < z < 1.4$, blueshifted \FeII\ absorption at greater than 100~\kms\ 
is seen toward 20\% of the population  (Martin \etal\ 2012).   Because the outflows are collimated, the ``outflow
fraction" must be signiifcantly higher than the ``blueshifted fraction".  
Some 2.5\% have velocities greater than 200~\kms, with the largest doppler components suggesting velocities 
up to 500~\kms.  The outflow fraction depends on the observed star-formation rate (SFR) and is three times higher 
for galaxies  with SFR $\approx 20-100~M_{\odot}~{\rm yr}^{-1}$ compared to those at $1-10~M_{\odot}~{\rm yr}^{-1}$. 
If these outflows persist at the observed rates for $\sim1$ Gyr, as Martin \etal\ (2012) suggest, they could remove a 
large mass of baryons, up to $10^{10}~M_{\odot}$.  Such metal-enriched outflows could be responsible for the large 
baryon reservoirs of gas and metals around star-forming galaxies observed in \OVI\  absorption as part of 
the COS-Halos project.  Tumlinson \etal\  (2011, 2013) found substantial  \OVI\ column densities, 
$N_{\rm OVI} \geq 10^{14.3}$~cm$^{-2}$ out to distances of 100-150 kpc,  toward a sub-sample of
28 galaxies with active star formation.  These \OVI-bearing halos surround galaxies with SFRs of
$1-10~M_{\odot}$~yr$^{-1}$ and specific SFRs  greater than $10^{-11}$~yr$^{-1}$.  

How far will these galactic winds expand before stalling?  Several simulations (Furlanetto \& Loeb 2003;  
Oppenheimer \& Dav\'e 2008) estimate that these outflows expand to  $\sim100$~kpc, distances inferred 
observationally in metal-line auto-correlation of \CIV\ absorbers (Martin \etal\ 2010).   A simple estimate illustrates 
the main effects, as winds encounter surrounding gas in the CGM and produce a ``galactopause" where the wind
ram pressure, $P_w = \rho_w V_w^2 = (\dot{M}_w V_w / \Omega_w  r^2)$, equals the confining gas pressure of
the CGM,  $P_{\rm CGM} = n_{\rm tot} \, k T = 2.17 n_H k T_{\rm CGM}$.  Here, 
$\dot{M}_w =  \Omega_w r^2 \rho_w V_w$ is the mass-loss rate of a conical outflow into total solid angle 
$\Omega_w$, $\rho_w$ is the mass density, 
and $V_w$ is the wind velocity.   Observers and modelers typically relate $\dot{M}_w $ to the SFR through a 
``mass-loading factor", defined here as $\beta_m = \dot{M}_w / {\rm SFR}$.  This mass loading has been inferred 
from optical and X-ray data to lie in the range $\beta_m = 1-3$ (Strickland \& Heckman 2009).    In a careful study 
of the dwarf starburst galaxy NGC~1569,  Martin, Kobulnicky, \& Heckman (2002) use observations of the SFR, 
H$\alpha$ emission, metallicities, and X-ray emission from the outflow to estimate
$M_X / M_{\rm ejecta} \approx 6-36$ for the mass ratio of X-ray emitting wind to stellar ejecta.  With standard 
assumptions about the ratio of ejecta mass to star formation, this range corresponds to mass loading factors
$\beta_m \approx 1-6$.  
  
For the CGM gas pressure in low-redshift, actively star-forming galaxies,  we use the observed \OVI\ column 
densities ($10^{14.5}$~\cd) and spatial impact parameters (100-150 kpc) from  COS-Halos data, which suggest 
total hydrogen column densities $N_H \approx (5-10)\times10^{18}$~\cd\ for solar oxygen abundances, (O/H) 
$\approx 5 \times10^{-4}$, and \OVI\ ionization fractions  $f_{\rm OVI} \approx 0.1-0.2$.  Stocke \etal\ (2013) 
estimated a mean pressure $\langle P/k \rangle  \approx 10$ cm$^{-3}$~K (with a range from 3 to 60 cm$^{-3}$~K) 
for warm clouds in the CGM of low-redshift galaxies.  Using \OVII\ X-ray absorption through the Galactic halo,
Miller \& Bregman (2013) modeled thermal pressures $P/k \approx [41, 24]$ cm$^{-3}$~K for halos with
characteristic scales of [50, 100] kpc and gas temperature $\log T = 6.1$.  

Thus, in the inner portions of the CGM that would initially confine the outflows, the gas pressure is scaled to  
$P_{\rm CGM} / k = 2.17 n_H T =  (40~{\rm cm}^{-3}~{\rm K}) P_{40}$, appropriate for hydrogen number density 
$n_H \approx 10^{-5}~{\rm cm}^{-3}$, halo virial temperature $T_{\rm CGM} \approx 2 \times10^6~{\rm K}$, and 
fully ionized gas with He/H $=  0.0833$ by number.  For wind speeds $V_w = (200~{\rm km~s}^{-1}) V_{200}$, 
the approximate radius of the galactopause is
\begin{equation}
   R_{\rm wind} = \left( \frac {\dot{M}_w V_w } {\Omega_w P_{\rm CGM} } \right)^{1/2} 
        = (140~{\rm kpc}) \left( \frac {{\rm SFR} }{10~M_{\odot}~{\rm yr}^{-1} } \right)^{1/2} \,
         \left( \frac {\Omega_w}{4 \pi} \right) ^{-1/2}  
         \left( \frac  {  \beta_m  \,  V_{200} } { P_{40} } \right) ^{1/2}   \;  .
\end{equation}    
The combination of parameters ($\beta_m \, V_{200} / P_{40}$) is expected to be of order unity, while 
bipolar wind solid angles $\Omega_w$ typically cover 10-40\% of $4\pi$ sterradians for starburst galaxies
(Veilleux, Cecil, \& Bland-Hawthorn 2005).  If the galaxy wind breaks through the higher-density CGM, the IGM
pressure will be considerably lower, and the metal-enriched winds will expand well beyond 200~kpc, depending on 
the duration of the starburst outflow.  Simulations of the low-redshift  IGM (e.g., Smith \etal\ 2011) find a wide range
of intergalactic gas pressures in the \Lya\ forest ($P/k \approx 0.1-2~{\rm cm}^{-3}~{\rm K}$)  and hotter \OVI-bearing 
gas  ($P/k \approx 3-30~{\rm cm}^{-3}~{\rm K}$).  

In the above estimate, the star-formation rates were scaled to $10~M_{\odot}~{\rm yr}^{-1}$.   In the sample of 
active star-forming galaxies at $z \sim 1$ (Martin \etal\  2012), the observed SFRs range from 20-100 
$M_{\odot}~{\rm yr}^{-1}$, while in the sample of 28 active COS-Halos galaxies at $z \approx 0.15-0.35$
(Tumlinson \etal\ 2013)  the SFRs are 1--10 $M_{\odot}~{\rm yr}^{-1}$.   These rates likely reflect the general 
decline in SFR history among all galaxies, from $z = 1-2$ to the present epoch.

\subsection{Local Environment:  Intervening Absorbers and Galaxies }

Indirect estimates of the sizes of extended gaseous halos and metal-enriched outflows around galaxies can
be made from observations of QSO absorption lines.  In a pioneering paper, Bahcall \& Spitzer (1969) proposed
that these absorbers are produced in ``extended halos of normal galaxies" and used the line frequency, $d{\cal N}/dz$, 
the number of absorbers per unit redshift, to constrain the halo cross section.  Their estimate, $R_0 \approx 100$~kpc, 
for the radius of a spherical halo is comparable to recent estimates from UV spectra of low-redshift QSO absorbers 
obtained from HST (Shull \etal\ 1996;  Stocke \etal\ 2006; Prochaska \etal\ 2011).  Those UV studies related line
frequency of absorbers in \HI\  (\Lya) or the \OVI\  doublet (1032, 1038~\AA) to  the absorber cross section and
space density, $\phi_{\rm gal}$, of galaxies, using a luminosity function to extrapolate to dwarf galaxy scales.   
In this formalism, the IGM absorbers are associated with the nearest-neighbor galaxy, a procedure expected to be
correct statistically, even though some absorbers have alternate possible galaxies within 200--600 kpc (Stocke \etal\
2013).  At redshifts $z \leq 0.4$ and ignoring cosmological effects, the number of of \OVI\ absorbers per unit redshift is
$d{\cal N} / dz \approx (c/H_0) ( \pi R_0^2) \phi_{\rm gal}$, from which one infers an absorber size 
\begin{equation}
    R_0 = \left[ \frac { d{\cal N}/dz } { \left( c / H_0 \right) \pi \, \phi_{\rm gal} } \right] ^{1/2} \approx
        \left( 157~{\rm kpc} \right) \left[ \frac {d{\cal N} / dz} {20} \right]^{1/2} 
        \left[  \frac {\phi_{\rm gal}} {0.06~{\rm Mpc}^{-3} }  \right] ^{-1/2}  \\  \; .
\end{equation}
This formula uses the observed $d {\cal N} / dz \approx  20$ down to 10~m\AA\  equivalent width for the \OVI\
1032~\AA\  absorption line in low-redshift absorber surveys (Danforth \& Shull 2008; Tilton \etal\ 2012).  We adopt
a Hubble constant $H_0 = (70~{\rm km~s}^{-1}~{\rm Mpc}^{-1}) h_{70}$, a Hubble length 
$(c/H_0) = (4286~{\rm Mpc}) \, h_{70}^{-1}$, and a galaxy luminosity function,
\begin{equation}
    \phi(L) \,  dL  = \frac {\phi_*}{L^*}   \left( \frac {L} {L^*} \right) ^{\alpha}  \exp (-L/L^*) \,  dL   \\   \;  , 
\end{equation} 
with normalization $\phi_* = (6.07 \pm 0.51)\times 10^{-3} \, h_{70}^3~{\rm Mpc}^{-3}$ at B-band absolute
magnitude $M^* = -20.37\pm0.04$ and faint-end slope $\alpha \approx -1.13 \pm 0.02$ given by
the Millennium galaxy catalog  (Driver \etal\ 2005) and converted to $h_{70} = 1$ scale.  The galaxy 
space density is scaled to 0.06~Mpc$^{-3}$, a value appropriate for extrapolation of $\phi(L)$ down 
to dwarf galaxies of luminosity $L \geq 0.1 L^*$, which have $\sim10$ times the space density of galaxies
with $L \geq L^*$ for $\alpha = -1.1\pm0.1$.  These sizes are consistent with surveys that attempt 
to match the nearest-neighbor galaxy, in 3D space, to low-redshift absorbers (Stocke \etal\ 2006, 2013).   
The galaxy spatial distributions in the surveys suggest that the nearest galaxies to \OVI\ absorbers lie at 
distances $\sim1$~Mpc from an $L^*$ galaxy and at $\sim200$~kpc from a galaxy at $0.1L^*$.

\subsection{ Halo Masses and Virial Radii from Galaxy Abundance Matching} 

Observers often lack direct measurements  of galaxy masses.  Instead, they estimate halo mass, $M_{\rm h}$, from the 
inferred stellar mass $M_*$ or galaxy luminosity $L$ (Shankar \etal\ 2006) using the technique of  ``abundance matching" 
(Behroozi \etal\ 2010; Moster \etal\ 2010, 2013).  By matching the abundance of dark-matter halos in simulations to observed 
galaxy distributions in stellar mass and luminosity, they attempt to characterize the relationship between the stellar 
masses of galaxies and the masses of the dark matter halos in which they live.   
Moster \etal\ (2010) obtain a parameterized stellar-to-halo mass relation by populating halos and subhalos in 
an N-body simulation with galaxies and requiring that the observed stellar mass function be reproduced.   In a 
different approach, Tacchella, Trenti, \& Carollo (TTC13) assume that the rest-frame UV luminosity and stellar mass 
of a galaxy are related to its dark-matter halo assembly and gas infall rate.  Galaxies are assumed to experience a burst 
of star formation at the halo assembly time (redshift $z = z_a$) followed by a constant star formation rate sustained by
steady gas accretion that provides the dominant contribution to the UV luminosity at all redshifts.  The model is 
calibrated by constructing a galaxy luminosity versus halo mass relation at $z = 4$ via abundance matching.
After this luminosity calibration, the model naturally fits the $z = 4$ stellar mass function and correctly predicts the
evolution of both luminosity and stellar mass functions from $z \approx 8$ to $z \approx 0$.   

Generically, halo-matching methods have many similarities, including the flattening at high $M_*$ of the relation 
that maps stellar mass $M_*$ into halo mass $M_h$.  The similar relation between luminosity $L$ of the central 
galaxy and the dark-matter halo has been explained (Cooray \& Milosavljevic 2005) as the result of a decline in 
merging efficiency for the accretion of satellites.  Flattening of the $M_*$ curve  at high $M_h$ predicts that
($z \approx 0$) galaxies with $M_* > 10^{11}~M_{\odot}$ reside in very massive halos, $M_h > 10^{13}~M_{\odot}$, 
although errors in $M_*$ (or $L$) can create large uncertainties in the inferred $M_h$.  A more subtle effect is that 
scatter in $M_*$ is amplified asymmetrically in $M_h$;  simply adopting the mean relation for $M_h(M_*)$ leads 
to errors (Behroozi \etal\ 2010).   Various halo-matching methods differ in specific details, including the redshift 
evolution of the ratio $M_* / M_h$.  Here, we adopt the TTC13 formalism, which has a clear physical basis for mass 
assembly and rest-frame UV luminosity evolution.

The COS-Halos study (Tumlinson \etal\ 2013;  Werk \etal\  2013) used the Moster \etal\ (2010) technique, whereas 
this paper adopts the TTC13 formalism.   Both methods find that galaxies with $L > L^*$ correspond to large halo 
masses.   However, as seen in Table~3, there are  differences between the two methods in the estimated halo 
masses and virial radii of the 44 galaxies in the COS-Halos sample.  The inferred values of $M_h$ and $R_{\rm vir}$ 
are particularly uncertain for galaxies with $L > L^*$.  Comparing values in Columns (5)--(8),  one sees that the new 
method gives  virial radii smaller by factors 0.4-0.8.  Statistically, the ratio of mean virial radii computed with the two 
methods is $\langle R_{\rm vir}^{\rm (new)} \rangle /  \langle R_{\rm vir}^{\rm (old)} \rangle  = 0.63 \pm 0.04 $,
with the largest change occurring for lower-mass halos, $\log M_* = 10.0-10.4$ (all masses are expressed 
in $M_{\odot}$ units).  Table 4 summarizes the statistical differences for the 44 galaxies in this sample, 
including subsets of galaxies at low, moderate, and high stellar masses.  

This decrease in $R_{\rm vir}$ occurs primarily because we evaluate the galaxy collapse and virialization at
overdensity, $\Delta_{\rm vir}$, at the redshifts of galaxy assembly, $z_a \approx 0.74-1.29$, rather than 
at the observed redshifts $z \approx 0.2-0.3$.  In addition, the conversion from $M_*$ to $M_h$ using modeling
or abundance matching differs between methods, particularly in its dependence on redshift.  Another difference 
is that we define $\Delta_{\rm vir}$, relative to the {\it matter} density rather than the {\it critical} density.  In a few 
instances with $L > L^*$ and large stellar masses, $\log M_* \geq 11.3$, we find large virial radii, 
$R_{\rm vir} > 400$~kpc, and anomalously large massive halos,  $\log M_h > 13.8$.  This result occurs because 
halo-matching algorithms and simulated luminosity functions find large stellar masses to be rare and associate 
them with very massive halos.  Applied to the large Local Group galaxies, halo-matching gives high values for the
Milky Way ($\log M_h = 12.55^{+0.18}_{-0.16}$ and $R_{\rm vir} = 153^{+25}_{-16}$~kpc for $\log M_* = 10.7\pm0.1$) 
and  M31 ($\log M_h = 13.35^{+0.24}_{-0.22}$ and $R_{\rm vir} = 303^{+69}_{-52}$~kpc for $\log M_* = 11.1\pm0.1$).  
Unfortunately for CGM studies, halo-matching for $L > L^*$ (high $M_*$) galaxies remains an uncertain technique, 
as demonstrated in Figure 1 of Stocke \etal\ (2013).

\section{CONCLUSIONS AND DISCUSSION} 

This paper reviews current observations and inferences about the spatial extent of galaxies, and the extent of 
their gravitational influence.  These included several dynamical measures:   the ``gravitational radius" 
$r_g = GM^2/ \vert W \vert$, the ``radius of influence" $R_{\rm infl} = GM / \sigma_{\parallel}^2$, and the 
``accretion radius"  $R_{\rm accr} = 2GM /c_s^2$.  
Clearly, these definitions are related for systems near dynamical equilibrium in the gravitational potential 
of the galactic halo.  For example, in Section 2.1 and Table~1, we found a ratio $r_g / R_{\rm vir} = 0.54-0.69$ for 
NFW halos with concentrations ranging from $c = 20$ to $c = 5$.   Various estimates of radial extent from 
rotation curves and injection of fast stars into extended halos  suggest 150-200~kpc extents for halo 
masses of  $10^{12}~M_{\odot}$.  Kinematic observations of the Milky Way and M31 give masses 
$\sim10^{12}~M_{\odot}$ within 60-80~kpc.    The inferred Galactic escape velocity,
$V_{\rm esc} = 533^{+54}_{-41}$~\kms, together with the local  circular velocity $V_c = 220$~\kms, is
consistent with halos extending beyond 80-100 kpc.  
Somewhat larger halo masses, $\log M_{\rm MW} = 12.2\pm0.1$ and $\log M_{\rm M31} = 12.3\pm0.1$
(all masses in $M_{\odot}$ units) have been inferred for the Milky Way and M31 out to larger radii, using 
globular clusters, satellite galaxies, and stellar streams.  Their mass sum, $\log M \approx 12.55\pm0.14$, 
is comparable to recent estimates of the timing mass of the Local Group, $\log M_{\rm LG} = 12.69^{+0.13}_{-0.19}$ 
(van der Marel 2012), although a recent study (Gonzalez \etal\ 2013) used cosmological simulations of 
halo pairs and a likelihood correction to find a lower value, $\log M_{\rm LG} = 12.38^{+0.09}_{-0.07}$.   For 
either of these masses, the Milky Way and Andromeda are likely to have radial extents  $\sim200$~kpc.

The answer to the question posed in the title to this paper depends on how one wishes to define or observe
the ``edge of a galaxy".  Does a galaxy end  because of gravity or outflows?   This paper suggests that the best 
definition of $R_{\rm vir}$ is in terms of gravitational binding.  Dark matter halos have very large extents, but 
galactic outflows can drive gas out of the stellar disk, into the (bound) halo, and often into the (unbound) CGM 
and IGM.   We have formulated a new definition of virial radius in this spirit, based on galaxy assembly in the 
past, when the background density was higher.  As a result, we find virial radii smaller by factors of 0.5-0.6, with 
important consequences for assessing whether the extended  absorption seen with COS is bound halo gas, 
or unbound gas on its way to the IGM. 

\newpage

\noindent
The main conclusions of this paper are as follows:  
\begin{itemize}

\item  A revised formulation of virial radius, $R_{\rm vir}(M_h, z_a)$, expresses the critical overdensity 
   $\Delta_{\rm vir}$ relative to the background matter density, ${\bar \rho}_m(z_a)$, evaluated at $z_a$, 
   the redshift of galaxy assembly when half the halo mass had collapsed.  Here, $z_a$ depends on $M_h$ 
   and is inferred from stellar mass $M_*$.   For dynamical consistency, the circumgalactic medium of large 
   galaxies is defined by their gravitationally bound regions, estimated from the virial radius or  
   gravitational radius,  $r_g(M_h, c) \approx (0.54-0.69)R_{\rm vir}$.
               
\item For galaxies with halo mass $11.0 \leq \log M_h \leq 14.0$ observed by HST/COS at redshifts
   $0.14 < z < 0.35$, the assembly redshifts are $1.35 > z_a > 0.81$.   Because virialization was
   largely determined by early collapse, when the background density was higher, we adopt 
   $\Delta_{\rm vir} \Omega_m \approx 200$ and $R_{\rm vir}$ is reduced by a factor $(1+z_a)$
   compared to $z \approx 0$.  
    
\item Halo masses can be estimated from stellar masses by halo-matching based on a physically motivated
  model (TTC13)  for galaxy assembly consistent with rest-frame UV  luminosity functions from $z = 0-8$.   
  Applied to 44 COS-Halos galaxies, this formalism yields virial radii smaller  by factors of 0.5-0.6.    
  Galaxies with $L \approx L^*$ ($\log M_h \approx 12.5$) have $R_{\rm vir} \approx 200$~kpc, 
  consistent with their observed radial extent. 
  
\item If confined by CGM gas pressure, the wind galactopause is predicted to occur at distances
  $R_{\rm wind} \approx 100-200$~kpc for SFR = 10-30 $M_{\odot}$~yr$^{-1}$, mass-loading factors
  $\beta_m \approx 1-3$, and the observed range of CGM gas pressures and wind velocities.  Stronger
  galaxy winds could break out into the lower density IGM, with radial extents beyond 200 kpc 
  determined by the strength, collimation, and duration of the outflow.  
  
\item For dynamical masses of the large Local Group galaxies, we adopt:  Milky Way ($\log M_h \approx 
  12.2\pm0.1$) and M31 ($\log M_h \approx 12.3\pm0.1$) with (gravitational) radial extents of 150-200 kpc.   
  In our new formalism, the Milky Way stellar mass, $\log M_* = 10.7 \pm 0.1$, would correspond to somewhat
  higher halo mass $\log M_h = 12.55^{+0.18}_{-0.16}$ and virial radiius $R_{\rm vir} = 153^{+25}_{-16}$~kpc 
  for assembly at $z_a =  1.06\pm0.03$.  The inferred $M_h$ for M31 is anomalously high
  ($\log M_h = 13.35^{+0.24}_{-0.22}$), which suggests caution in using halo-matching to obtain $M_h$ and
  $R_{\rm vir}$ for massive, luminous galaxies ($L > L^*$).  

\end{itemize} 

As noted earlier, the new definition of $R_{\rm vir}$ is important in deciding whether gas in the vicinity 
of galaxies is gravitationally bound or unbound.   Whether galaxies are ``open or closed boxes" helps to
determine the metal evolution of galaxies and the extent of metal-pollution of the IGM.  It also suggests that 
astronomers should be careful in making interchangeable use of the terms CGM and Halo.  Matter in 
the galactic halo is {\it gravitationally bound}, while the circumgalactic (gaseous) medium may in fact be
 {\it unbound outflows} that will merge into the IGM.   What many observers are calling the CGM is probably
gas  at the  CGM--IGM interface.    

In observations of large metal-enriched reservoirs around star-forming galaxies (Tumlinson \etal\ 2013;  
Stocke \etal\ 2013), the radial offset distances $R$ of the QSO sight lines were normalized to $R_{\rm vir}$, 
with much of the metal-enriched gas within $R < R_{\rm vir}$.  With the revised (smaller) virial radii (Table~3 
and Figure~1) some of this metal-enriched gas extends beyond the region of strong gravitational influence, 
probably on its way out to the IGM.  This trend is seen in the 28 actively star-forming galaxies in the 
COS-Halos sample, whereas the undetected \OVI\ around 16 passive galaxies all had sight lines with 
$R < R_{\rm vir}$.   In future work, both data sets will be examined carefully with the new formalism.


\acknowledgments

\noindent
I thank Crystal Martin, Michele Trenti, Massimo Ricotti, John Stocke, Brian Keeney, Blair Savage, and Vasily 
Belokurov for helpful discussions, and Evan Tilton and Joshua Moloney for comments on the manuscript.  
This research was supported by the Astrophysical Theory Program  at the University of Colorado Boulder 
(grant NNX07-AG77G from NASA).   I am grateful to the Institute of Astronomy at Cambridge University
for their stimulating scientific environment and support through the Sackler Visitor Program.


\newpage


\begin{deluxetable}{lccc}
\tabletypesize{\footnotesize}
\tablecaption{\bf Gravitational Radius ($r_g$) for Various Density Models}  
\tablecolumns{4}
\tablewidth{0pt}
\tablehead{   \colhead{Model}              &   \colhead{Density $\rho(r)$\tablenotemark{a}}   &  
     \colhead{Scale\tablenotemark{a}}  &  \colhead{$r_g$\tablenotemark{a} }
   }
\startdata
Uniform Sphere            &  $\rho = \rho_0~{\rm for}~r \leq R$    &     $R$     &     $1.67R$    \\
Plummer Model            &  $(3M/4 \pi b^3)[1 + (r/b)^2]^{-5/2}$  &      $b$     &     $3.40b$    \\
Jaffe Model                    & $\rho_0 (r/a)^{-2}(1+r/a)^{-2}$           &      $a$     &     $2a$          \\
Hernquist Model           & $\rho_0 (r/a)^{-1}(1+r/a)^{-3}$           &      $a$     &     $6a$          \\
NFW Model  ($c=5$)    & $\rho_0 (r/r_s)^{-1}(1+r/r_s)^{-2}$    &      $r_s$  &     $3.44r_s$     \\
NFW Model  ($c=10$)  & $\rho_0 (r/r_s)^{-1}(1+r/r_s)^{-2}$    &      $r_s$  &     $6.41r_s$      \\
NFW Model  ($c=15$)  & $\rho_0 (r/r_s)^{-1}(1+r/r_s)^{-2}$    &      $r_s$  &     $8.81r_s$      \\
NFW Model  ($c=20$)  & $\rho_0 (r/r_s)^{-1}(1+r/r_s)^{-2}$    &      $r_s$  &     $10.8r_s$      \\
\enddata

\tablenotetext{a}{Radial density distribution, characteristic radial scale, and gravitational radius, 
$r_g = GM^2/ \vert W \vert$, defined in terms of mass and potential energy of system 
(Binney \& Tremaine 2008).  For the  NFW density distribution, the total mass diverges logarithmically 
with radius, but we define $r_g$ for the mass inside $R_{\rm vir} = c \, r _s$ as shown in Equation (9).  
}
\end{deluxetable}



\begin{deluxetable}{cccccc}
\tabletypesize{\footnotesize}
\tablecaption{\bf Redshifts of Galaxy Halo Assembly ($z_a$)  }  
\tablecolumns{6}
\tablewidth{0pt}
\tablehead{   \colhead { $\log (M_h/ M_{\odot})$ }        &   \colhead { $z_a$\tablenotemark{a} }  &
   \colhead{}   &  \colhead{} &   \colhead { $\log (M_h/ M_{\odot})$ }    &   \colhead { $z_a$\tablenotemark{a} } 
}
                                                                                     
\startdata
9.00    &   1.709    &    &    &   12.25   &   1.112   \\
9.25    &   1.666    &    &    &    12.50   &   1.070  \\
9.50    &   1.620    &    &    &    12.75   &   1.026  \\
9.75    &   1.572    &    &    &    13.00   &   0.979  \\
10.00   &  1.531    &    &    &   13.25   &   0.936  \\
10.25  &   1.486    &    &    &   13.50   &   0.891  \\
10.50   &   1.439   &    &    &   13.75   &   0.844  \\
10.75   &   1.394   &    &    &   14.00   &   0.806  \\
11.00   &   1.346   &    &    &   14.25   &   0.765  \\
11.25   &   1.302   &   &     &   14.50   &   0.722  \\
11.50   &   1.255   &   &     &   14.75   &  0.684   \\
11.75   &   1.216   &   &     &   15.00   &   0.645  \\
12.00   &   1.155   &   &   
     
\enddata

\tablenotetext{a} {Halo assembly formalism (TTC13) model) assumes that half 
the halo mass ($M_h/2$) is assembled at redshift $z_a$ (Trenti, Perna, \& Tacchella
2013).  Values of $z_a$ are used in virial radius (Equation 12).
See Sections 2.3 and 3.4  for discussion and applications.  }
   
\end{deluxetable}



\begin{deluxetable}{lccccccrrr}
 \tabletypesize{\footnotesize}
\tablecaption{\bf  Masses and Virial Radii of 44 COS-Halos Galaxies\tablenotemark{a} }  
\tablecolumns{10}
\tablewidth{0pt}
\tablehead{  \colhead{QSO~(Galaxy)\tablenotemark{a} } &  \colhead{$z$\tablenotemark{a} }               &
       \colhead{$(L/L^*)$\tablenotemark{a} }                        &  \colhead{$\log M_*$\tablenotemark{a} }   &
       \colhead{$\log M_h$\tablenotemark{b} }            &  \colhead{$\log M_h$\tablenotemark{b} }          &
       \colhead{$R_{\rm vir}$ \tablenotemark{b} }        &  \colhead{$R_{\rm vir}$\tablenotemark{b} }       & 
       \colhead{$z_a$\tablenotemark{b} }        & \colhead{$R$\tablenotemark{a} }       \\
       &   &   &  & (old) &  (new)  & (old)  &  (new) &   \\
        (1)   &  (2)  &  (3)  &  (4)  &  (5)  &  (6)  &  (7)  &  (8)  &  (9)  & (10)  
     }
\startdata
J0226+0015(268,22)  & 0.22744 & 0.51 & 10.8 & 12.62 & 12.73  & 303 & 178   & 1.03   &   80 \\
J0401-0540~~(67,24) & 0.21969 & 0.37 & 10.2 & 12.08 & 11.86  & 200 & 85     & 1.19   &   86 \\
J0803+4332(306,20)  & 0.25347 & 1.63 & 11.3 & 13.48 & 13.84  & 581 &  462  & 0.83   &   79 \\
J0910+1014~(34,46)  & 0.14274 & 0.60 & 10.6 & 12.48 & 12.39  & 279 & 133   &  1.09  &   116 \\ 
J0910+1014(242,34)  & 0.26412 & 2.16 & 11.5 & 13.76 & 14.40  & 716 & 747   &  0.74  &   139 \\ 
J0914+2823~(41,27)  & 0.24431 & 0.37 & 9.8   &  11.87 & 11.48 & 169 & 61      & 1.26   &   104 \\
J0925+4004(196,22)  & 0.24745 & 1.48 & 11.3 & 13.45 & 13.84  & 569 & 462   & 0.83   &   86   \\
J0928+6025(110,35)  & 0.15400 & 0.52 & 10.8 & 12.65 & 12.73  & 317 & 178   & 1.03   &    94  \\
J0935+0204~(15,28)  & 0.26228 & 0.79 & 11.0 & 12.88 & 13.13  & 365 & 251   & 0.95   &    114 \\
J0943+0531(106,34) & 0.22839  & 0.74 & 10.8 & 12.61 & 12.73  & 300 & 178   & 1.03   &    125 \\
J0943+0531(216,61) & 0.14311  & 0.72 & 11.0 & 12.80 & 13.13  & 382 & 251   & 0.95   &    154 \\
J0943+0531(227,19) & 0.35295  & 0.24 & 9.6   & 11.66 & 11.31  & 141  &  69     & 1.29   &    95 \\
J0950+4831(177,27) & 0.21194  & 1.44 & 11.2 & 13.30 & 13.59  & 511  & 371   & 0.88   &    94 \\
J1009+0713(204,17) & 0.22784  & 0.28 & 9.9   & 11.90  & 11.56  & 174  & 66     & 1.24   &   62  \\
J1009+0713(170,9)   & 0.35569  & 0.29 & 10.3  & 12.06 &  11.98 &  189 & 94     & 1.16   &  45  \\
J1016+4706(274,6)   & 0.25195  & 0.23 & 10.2  & 12.10 &  11.86 & 202  & 85   & 1.19   & 24  \\
J1016+4706(359,16) & 0.16614  & 0.44 & 10.5  & 12.35 & 12.24  & 251   & 117  & 1.12   & 46  \\
J1112+3539(236,14) & 0.24670  & 0.52 & 10.3  & 12.17 & 11.98  & 214   & 94    & 1.16   & 54 \\
J1133+0327(110,5)   & 0.23670  & 1.94 & 11.2  & 13.19 & 13.59  & 515  & 371   & 0.88    & 14 \\
J1133+0327(164,21) & 0.15449  & 0.19 & 10.1  & 12.09 & 11.76  & 206  & 78     & 1.21    & 56 \\
J1157-0022(230,7)    & 0.16378  & 0.55 & 10.9  & 12.72 & 12.91  & 334  & 207   & 1.00   &  20 \\
J1220+3853(225,38) & 0.27371 & 0.66 & 10.8   & 12.53 & 12.73  & 279  & 178   & 1.03   & 159 \\
J1233+4758(94,38)   & 0.22210 & 0.70 & 10.8   & 12.58  & 12.73  & 295 & 178   & 1.03   & 137 \\
J1233-0031(168,7)    & 0.31850 & 0.37 & 10.6    & 12.30  & 12.39 & 230 & 133   & 1.09   & 33  \\ 
J1241+5721(199,6)  & 0.20526  & 0.20 & 10.2   & 12.10   & 11.86 & 205 &  85    & 1.19    & 21 \\
J1241+5721(208,27) & 0.21780 & 0.20 & 10.1   & 12.02   & 11.76 & 192 &  78    & 1.21    & 92  \\
J1245+3356(236,36) & 0.19248 & 0.20 & 9.9     & 11.81   & 11.56 &  178 &  66   & 1.24  & 117  \\
J1322+4645(349,11) & 0.21418 & 0.58 & 10.8   & 12.50   & 12.73  & 303 &  178 & 1.03  & 39  \\
J1330+2813(289,28) & 0.19236 & 0.24 & 10.3   & 12.22   & 11.98  & 225 &  94   & 1.16    & 91  \\
J1342-0053(157,10)  & 0.22702  & 1.08 & 11.0  &  12.67  & 13.13 & 345 & 251   & 0.95 & 37    \\
J1342-0053(77,10)    & 0.20127  & 0.25  & 10.5  &  12.34  & 12.24  & 247 & 117  & 1.12 &  34    \\
J1419+4207(132,30) & 0.17925 & 0.55  & 10.6 & 12.46    & 12.39  & 272  & 133  & 1.09  & 92 \\
J1435+3604(126,21) & 0.26226 & 0.33  & 10.4 & 12.20    & 12.01  & 218  &  97   & 1.15  & 86   \\
J1435+3604(68,12)    & 0.20237 & 1.37  &  11.1 & 13.08   & 13.35   & 433  & 302  & 0.92  &  40  \\
J1437+5045(317,38) & 0.24600 & 0.46  & 10.2  & 12.06   & 11.86  & 196  &  85    & 1.19 &  149 \\
J1445+3428(232,33) & 0.21764 & 0.29  & 10.4  & 12.26   & 12.01  & 230  & 97     & 1.15  &   117   \\
J1514+3619(287,14) & 0.21223 & 0.18  & 9.7    &  11.81  & 11.39   &164  & 57     & 1.28  &   49   \\
J1550+4001(197,23) & 0.31247 &  1.78 & 11.4  & 13.50  & 14.10   & 578 & 577   & 0.79   &   106   \\
J1550+4001(97,33)   & 0.32179 &  0.82 & 10.9  & 12.54   & 12.91  & 311 & 207    & 1.00  &   155  \\
J1555+3628(88,11)   & 0.18930 &  0.54 & 10.5  &  12.38  & 12.24  & 254  & 117  & 1.12   & 35    \\
J1617+0638(253,39) & 0.15258 & 2.65  & 11.5  & 14.03  & 14.40  &  912 & 747  & 0.74    &   104 \\
J1619+3342(113,40) & 0.14137 & 0.19  & 10.1  &12.12   & 11.76  &  211 & 78   & 1.21     &   100 \\
J2257+1340(270,40) & 0.17675 & 0.67  & 10.9   & 12.78 & 12.91  &  348 & 207  & 1.00    &   120 \\
J2345-0059(356,12) & 0.25389 & 0.80  & 10.9  &12.63   & 12.91  &  304 & 207   & 1.00    &   48 \\

\enddata
 
\tablenotetext{a}{COS-Halos survey data on 44 galaxies (Tumlinson \etal\ 2013; Werk \etal\ 2013).
   Column 1 lists background QSO and intervening galaxy, following their notation (position angle 
   with respect to QSO and angular separation in arcsec).  Columns 2-4 give galaxy redshift, luminosity 
   in $L^*$, and stellar mass $M_*$ computed from SDSS {\it ugriz} photometry (Werk \etal\ 2012) 
   updated  by Tumlinson \etal\ (2013).   Column (10) gives the impact parameter $R$ (in kpc) of the
   QSO sight line relative to galaxy, using angular separation converted to angular-size distance using 
   WMAP-9 $\Lambda$CDM cosmological parameters ($h=0.7$,  $\Omega_m = 0.275$, 
   $\Omega_{\Lambda} = 0.725$) that differ slightly from values in Tumlinson \etal\ (2013).  }

\tablenotetext{b}{ Galaxy halo masses $M_h$ (in $M_{\odot}$) were estimated two ways:  
  Column 5 (old) by abundance-matching (Moster \etal\ 2010, as used by Tumlinson \etal\ 2013);   
  Column 6 (new) by modeling luminosity function and halo matching (Tacchella \etal\ 2013).  
  Virial radii (in kpc) were also estimated two ways:  Column~7 (Tumlinson \etal\ 2013) and 
  Column~8 (this paper eq.\ [12]) using new halo mass $M_h$ from Column 6 and halo assembly 
  redshift $z_{\rm vir} = z_a$ (column 9). }

\end{deluxetable}



\begin{deluxetable}{lccccccc}
\tabletypesize{\footnotesize}
\tablecaption{\bf Statistical Averages\tablenotemark{a} for COS-Halos Sample  }  
\tablecolumns{8}
\tablewidth{0pt}
\tablehead{   \colhead{Sample}   &  $N_{\rm gal}$  &   \colhead{ $\langle \log M_* \rangle$ }   & 
     \colhead{ $\langle \log M_h^{{ \rm (old)}} \rangle$ }   &   
     \colhead{ $\langle \log M_h^{{ \rm (new)}} \rangle$ }   &
     \colhead{$\langle R_{\rm vir}^{ {\rm (old)}}\rangle$ }  &  
     \colhead{$\langle R_{\rm vir}^{ {\rm (new)}} \rangle$ } & 
     \colhead{$\langle R_{\rm vir}^{ {\rm (new)}} / R_{\rm vir}^{{\rm (old)} } \rangle$ }   \\
  &  & ($M_*$ in $M_{\odot}$)   &  ($M_h$ in $M_{\odot}$)  &  ($M_h$ in $M_{\odot}$)   &   (kpc)   &   (kpc)   &      
   }
\startdata

All Galaxies                      &  44  &  10.61   &  12.53   &  12.55   &  315  &  200  &  0.563 \\
$\log M_* = 10.1-10.4$  &  12   & 10.23   &  12.12   &  11.89   &  207  &   81   &  0.422  \\
$\log M_* = 10.8-10.9$  &  10  &  10.84   &  12.62   & 12.80    &  309  &  190  &  0.613  \\
$\log M_* = 11.2-11.5$  &   7   &   11.34  &  13.53   &  13.97   &  626  &  534  &  0.845    \\

\enddata

\tablenotetext{a}{Statistical means of halo masses ($\log M_h$) and virial radii ($R_{\rm vir})$ 
computed for the two different methods, labeled {\it old} and {\it new}, from data in Table 3.  
Values are shown for all 44 COS-Halos galaxies and for three subsets at low, medium, and high 
stellar masses ($M_*$).   The last column shows averages of the ratio of $R_{\rm vir}$ in the 
two methods. }

\end{deluxetable}


\clearpage


\begin{figure}[ht]
  \plotone{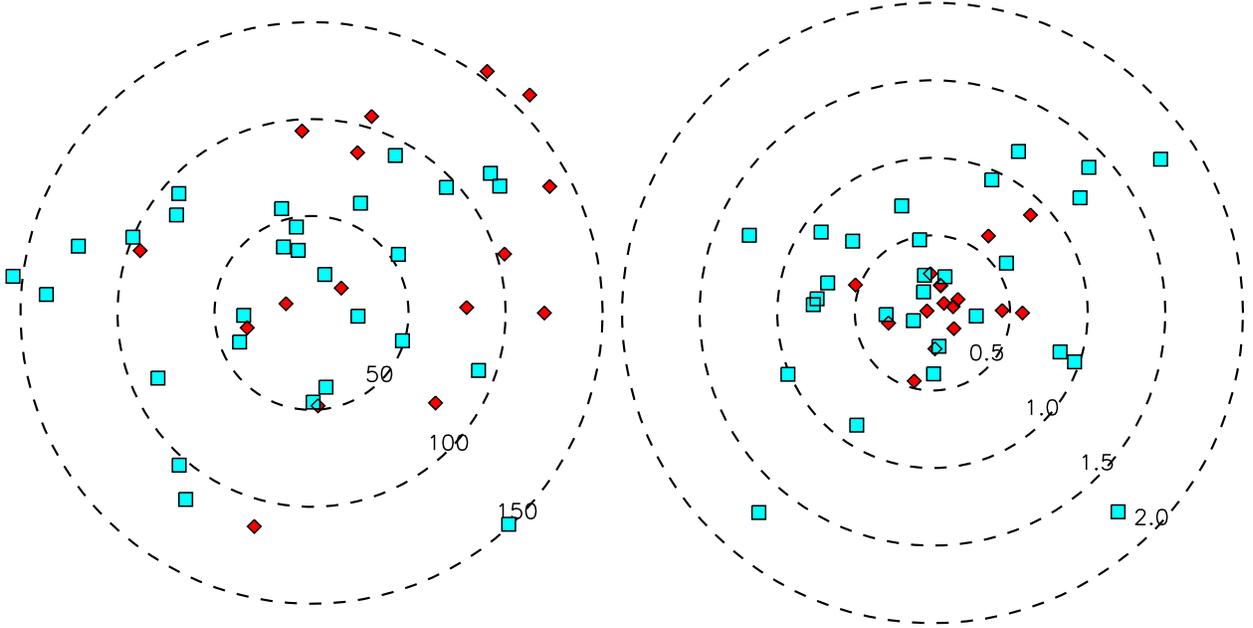}  
  \caption{Distribution of  COS-Halos targets (Tumlinson \etal\ 2013).  {\it Left Panel}: Impact parameter 
   ($R$ in kpc) and position angle (measured from N to E) of the background AGN relative to 44 galaxies in 
   survey.  Blue squares show 28 actively star-forming galaxies, and red diamonds show 16 passive galaxies.  
   These impact parameters have been slightly updated using WMAP-9 cosmological parameters (see text). 
  {\it Right Panel}:  Normalized radial extent,  $R / R_{\rm vir}$, relative to the newly computed virial radii
  (Table 3).   Many of the sight lines through star-forming galaxies probe gas outside $R_{\rm vir}$,
   while  {\it all} of the passive galaxies probe gas at $R < R_{\rm vir}$.  This dichotomy suggests the role
   of feedback from strong galactic outflows in metal transport to the CGM and IGM.  
   }
  
  \end{figure}


 \end{document}